\documentclass[a4paper,UKenglish,autoref,11pt]{article}

\usepackage{microtype}

\usepackage{soul}

\title{On Hardness of Testing Equivalence to Sparse Polynomials Under Shifts}

\author{
  Suryajith Chillara\thanks{Part of this work was done while the author was visiting Tel Aviv University, hosted by Amir Shpilka.}~~\href{mailto:suryajith.chillara@iiit.ac.in}{\faEnvelopeO}\\ IIIT-Hyderabad, India.
  \and
  Coral Grichener~\href{mailto: coralgri@gmail.com }{\faEnvelopeO}\\ Google, Israel.
  \and
  Amir Shpilka\thanks{The research leading to these results received funding from the Israel Science Foundation (grant number 514/20) and from the Len Blavatnik and the Blavatnik Family foundation.}~~\href{mailto:shpilka@tauex.tau.ac.il}{\faEnvelopeO}\\ Tel Aviv University, Israel.
}

\usepackage[dvipsnames,svgnames,x11names,hyperref, table]{xcolor}
\usepackage[framemethod=TikZ]{mdframed}
\usepackage{tikz,epigraph,verbatim,thmtools,thm-restate,ccicons,setspace,microtype}
\usepackage{array}
\usepackage{amsmath}
\usepackage{amssymb}
\usepackage{amsthm}
\usepackage{fullpage}
\usepackage{caption}

\usepackage{todonotes}

\usepackage{libertine}
\usepackage{eulervm}

\usepackage[T1]{fontenc}
\usepackage{inconsolata}

\usepackage{fontawesome}

\definecolor{bloodRed}{RGB}{92,3,37}
\definecolor{slightPurple}{RGB}{166,6,66}
\definecolor{slighterPurple}{RGB}{184,43,87}
\usepackage[colorlinks,backref,pagebackref]{hyperref}
\hypersetup{
  linkcolor=slightPurple, 
  citecolor=slighterPurple, 
  urlcolor= bloodRed 
}

\allowdisplaybreaks
\usepackage{multicol}
\usepackage[ruled,vlined, linesnumbered]{algorithm2e}
\usepackage[capitalize, nameinlink, noabbrev]{cleveref}

\crefname{algocf}{alg.}{algs.}
\Crefname{algocf}{Algorithm}{Algorithms}

\newtheorem{theorem}{Theorem}
\newtheorem{corollary}[theorem]{Corollary}
\newtheorem{lemma}[theorem]{Lemma}

\newtheorem{remark}[theorem]{Remark}

\newcommand{\HN}{\mathsf{HN}}
\newcommand{\PolySparse}{\mathsf{SparseShift}}
\newcommand{\PolyProj}{\mathsf{PolyProj}}
\newcommand{\ShiftEquiv}{\mathsf{ShiftEquiv}}

\newcommand{\maxthreelin}{\text{Max-3Lin}}

\newcommand{\inparen }[1]{\left(#1\right)}             
\newcommand{\inbrace }[1]{\left\{#1\right\}}           

\newcommand{\abs}[1]{\left|#1\right|}                  

\newcommand{\F}{\mathbb{F}}
\newcommand{\N}{\mathbb{N}}
\newcommand{\Q}{\mathbb{Q}}
\newcommand{\Z}{\mathbb{Z}}
\newcommand{\R}{\mathbb{R}}
\newcommand{\C}{\mathbb{C}}

\newcommand{\poly}{\operatorname{poly}}

\newcommand{\veca}{\mathbf{a}}
\newcommand{\vecb}{\mathbf{b}}
\newcommand{\vecc}{\mathbf{c}}

\newcommand{\vece}{\mathbf{e}}

\newcommand{\NP}{\mathsf{NP}}

\renewcommand{\epsilon}{\varepsilon}

\renewcommand{\epsilon}{\varepsilon}
\newcommand{\ignore}[1]{}

\makeatletter

\begin{document}

\maketitle

\begin{abstract}

  We say that two given polynomials $f, g \in R[x_1, \ldots, x_n]$, over a ring $R$, are equivalent under shifts if there exists a vector $(a_1, \ldots, a_n)\in R^n$ such that $f(x_1+a_1, \ldots, x_n+a_n) = g(x_1, \ldots, x_n)$. This is a special variant of the polynomial projection problem in Algebraic Complexity Theory. 
 
  Grigoriev and Karpinski (FOCS 1990), Lakshman and Saunders (SIAM J. Computing, 1995), and Grigoriev and Lakshman (ISSAC 1995) studied the problem of testing polynomial equivalence of a given polynomial to \emph{any} $t$-sparse polynomial, over the rational numbers, and gave exponential time algorithms. In this paper, we provide  hardness results for this problem.
  
  Formally, for a ring $R$, let $\PolySparse_R$ be the following decision problem -- Given a polynomial $P(X)$, is there a vector $\veca$ such that $P(X+\veca)$ contains fewer monomials than $P(X)$. We show that $\PolySparse_R$ is at least as hard as checking if a given system of polynomial equations over $R[x_1,\ldots, x_n]$ has a solution (Hilbert's Nullstellensatz). 
  As a consequence of this reduction, we get the following results.
  \begin{enumerate}
      \item $\PolySparse_\mathbb{Z}$ is undecidable.
      \item For any ring $R$ (which is not a field) such that $\HN_R$ is $\NP_R$-complete over the Blum-Shub-Smale model of computation, $\PolySparse_{R}$ is also $NP_{R}$-complete.  In particular, $\PolySparse_{\Z}$ is also $\NP_{{\Z}}$-complete.
  \end{enumerate}

  We also study the gap version of the $\PolySparse_R$ and show the following. 
  \begin{enumerate}
      \item For every function $\beta:\N\to\R_+$ such that $\beta\in o(1)$, $N^\beta$-gap-$\PolySparse_\mathbb{Z}$ is also undecidable (where $N$ is the input length).
      \item For $R=\F_p, \mathbb{Q}, \mathbb{R}$ or $\mathbb{Z}_q$ and for every $\beta>1$  the $\beta$-gap-$\PolySparse_R$ problem is NP-hard. Furthermore, there exists a constant $\alpha>1$ such that for every $d = O(1)$ in the sparse representation model, and for every $d\leq n^{O(1)}$ in the arithmetic circuit model, the $\alpha^d$-gap-$\PolySparse_R$  problem is  NP-hard when given polynomials of degree at most $d$, in $O(nd)$ many variables, as input. 
  \end{enumerate}
\end{abstract}
\newpage
\section{Introduction}

This paper studies the following question: given an $n$-variate polynomial $f(X)$, over a ring $R$\footnote{From now on, $R$ always denotes an integral domain, i.e. a commutative ring with a unit, which is also a domain, and $\F$ a field ($\Q,\R$ and $\C$ are, as usual, the rational, real and complex fields, respectively).}, how difficult is the task of finding a shift $\vecb\in R^n$ such that $f(X +\vecb)$ has fewer monomials than $f$. 

Before proceeding we would like to discuss the issue of representation of polynomials. There are several natural settings -- representation as vector of coefficients or as arithmetic circuits -- and two different models -- the white-box and black-box models. The most obvious representation is the \emph{dense representation} in which $n$-variate polynomials of degree $d$ are represented as a vectors of coefficients of length ${n+d\choose d}$. In this setting we assume that the vector is given as input to the algorithm. A more concise representation is the \emph{sparse representation} in which a polynomial is represented as a list of pairs of exponent vectors and coefficients. In the black-box setting we only assume that the algorithm has black-box access to the polynomial  (though the important parameters such as number of variables and degree are known to the algorithm). I.e., the algorithm is restricted to asking the polynomial for its values on different inputs. 
Another natural model is representing polynomials as arithmetic circuits. That is, the algorithm will get as input an arithmetic circuit computing the polynomial. In the white-box setting the algorithm is explicitly given the circuit so it has access to the graph of computation etc. In the black-box model the algorithm only has black-box access to the circuit (though the important parameters such as size, depth, number of variables etc. are known to the algorithm).

One of the most important questions in the area of Algebraic Complexity Theory is the problem of checking if two polynomials are equivalent under affine transformations. In generality this problem is also called the polynomial projection problem. Ignoring issues of representations the problem is the following.
\begin{center}
  \begin{tabular}{|m{15cm}|}
    \hline
    \vspace{0.1cm}
    {\large Polynomial Projection ($\PolyProj_\F$):}
    \vspace{0.1cm}\\
    \hline
    \vspace{0.1cm}
    Given two polynomials $f\in \F[y_1, \ldots, y_m]$ and $g\in \F[x_1,\ldots, x_n]$, over a field $\F$, output an $m\times n$ matrix $A$ and a vector $\vecb\in \F^{m}$ such that $ g(x_1, \ldots, x_n) = f(A\cdot \begin{bmatrix}x_1 & x_2 &\ldots & x_n\end{bmatrix}^T+\vecb)$ if such a pair exists, or output ``FAIL'' otherwise.\\
    \hline
  \end{tabular}
\end{center}

For example, the holy grail of algebraic complexity, Valiant's Extended Hypothesis  is an instance of the polynomial projection problem. Recall that the hypothesis says that the permanent of an $n\times n$ matrix cannot be represented as a polynomial projection of determinant of any $m\times m$ matrix, for any  $m$ that is polynomial in $n$~\cite{v79}. Kayal~\cite{Kay12} showed that the problem of polynomial projection is $\NP$-hard in general. However, for specific instances of the polynomial  $g$, under the requirement that the matrix $A$ has full rank (or that it is random), Kayal~\cite{Kay12} gave efficient randomized algorithms in the black-box model (i.e. assuming only black-box access to $f$).

Since studying polynomial equivalence under such projections is $\NP$-hard in general, the following \emph{simpler} question was considered.
\begin{center}
  \begin{tabular}{|m{15cm}|}
    \hline
    \vspace{0.1cm}
    {\large Polynomial Equivalence under Shifts ($\ShiftEquiv_\F$):}
    \vspace{0.1cm}\\
    \hline
    \vspace{0.1cm}
    Given two polynomials $f,g\in \F[x_1, \ldots, x_n]$ output a vector $(b_1, \ldots, b_n)\in \F^{n}$ such that $ g(x_1, \ldots, x_n) = f(x_1+b_1, \ldots, x_n+b_n)$ if such a vector exists, or output ``FAIL'' otherwise.\\
    \hline
  \end{tabular}
\end{center}
To the best of our knowledge the notion of studying polynomial equivalence under shifts first appeared in \cite{GK93} and it was formally addressed by Grigoriev in \cite{G97}. For polynomials of degree $d$ over $n$ variables, Grigoriev~\cite{G97} gave a deterministic algorithm over fields of zero characteristic, a randomized algorithm over prime residue fields, and a quantum algorithm over fields of characteristic $2$, all of which run in time polynomial in the dense representation. That is, the running time is polynomial in ${n+d \choose d}$. If the degree of the polynomial grows as a function of the number of variables or vice versa,  the algorithms presented by Grigoriev require \emph{exponential} time in the number of variables, even if the polynomial can be represented by a small arithmetic circuit or if it has polynomially many monomials. It is a natural question to ask if the complexity of the algorithms can be brought down when the input to the algorithm is provided in some succinct representation -- for example, as an arithmetic circuit. In such a setting, Dvir, Oliveira and Shpilka~\cite{DOS14} showed that given just a black box access to the polynomials $f$ and $g$ on $n$ variables, and given a bound on the degree $d$ and circuit size $s$, there is a randomized algorithm that runs in time $\poly(n,d,s)$ and solves the polynomial equivalence under shifts problem. 
The randomness in their algorithm only stems from polynomial identity testing  (PIT), which is a sub-routine of their algorithm, and hence 
equivalence under shifts in this setting can be derandomized if and only if PIT can be derandomized (clearly PIT is a special case of equivalence under shifts when $g$ is the zero polynomial).

A polynomial $f$ is said to be $t$-sparse if the number of monomials with non-zero coefficients in $f$ is at most $t$.
In the literature, an  $n$ variate polynomial is generally said to be sparse if the  number of monomials in it with non-zero coefficients is at most $\poly(n)$. Equivalently, a sparse polynomial is a polynomial that can be computed by a depth two $\Sigma\Pi$ arithmetic circuit with a polynomial bound on the top fan-in. 
Sparse polynomials are extremely well studied because of their simplicity and as a result many efficient algorithmic results are known for them~\cite{BT88, KLW90, BT91, CDGK91, GKS90, GK91, GK93, LS95, SY11}. 

A variant of the polynomial projection problem asks if a given polynomial is equivalent to a sparse polynomial under affine transformations. This can be seen as a variant of the classical Minimum Circuit Size Problem (MCSP) where given the truth table of a function we wish to find the minimal circuit computing it. In this case the circuit we are seeking is a very structured $\Sigma\Pi\Sigma$ circuit that is obtained by composing a $\Sigma\Pi$ circuit with an affine transformation. As this set of polynomials is dense inside the class $\Sigma\Pi\Sigma$ it is an interesting family to study (see \cite{MediniS21}).  Grigoriev and Karpinski~\cite{GK93} were the first to consider this variant of the polynomial projection problem. Specifically, they studied the following problem (in the dense representation model) -- given a polynomial $P(X)$, over the rationals, and a parameter $t$ output a matrix $A$ and a vector $\vecb$, if they exist, such that the polynomial $P(A\cdot X+\vecb)$ has at most $t$ monomials. They gave an algorithm whose complexity is $O(M\cdot d^{n^4})$ where $M$ is a bound on the size of coefficients of the input polynomial. Lakshman and Saunders~\cite{LS95} considered the problem of testing the equivalence of univariate polynomials (over $\mathbb{Q}$) to $t$-sparse polynomials under just shifts instead of affine linear transformations. They provided sufficient conditions for uniqueness and rationality of a $t$-sparsifying shift. Grigoriev and Lakshman~\cite{GL00} extended these criterion to multivariate polynomials. They also gave algorithms for polynomials with finitely many sparsifying shifts\footnote{Over $\mathbb{Z}, \mathbb{Q}, \mathbb{R}, \mathbb{C}$ etc., it may happen that there are infinitely many $t$-sparsifying shifts for a given polynomial. Grigoriev and Lakshman~\cite{GL00} give algorithms for polynomials that are guaranteed to have finitely many $t$-sparsifying shifts.} that run in deterministic time $(dt)^{O(n)}$ and randomized time $t^{O(n)}$. In the past two decades, these exponential time algorithms could not be improved and this is a major motivation behind our study of hardness of this problem. We state the following more general problem to allow polynomials over rings. 

\begin{center}
  \begin{tabular}{|m{15cm}|}
    \hline
    \vspace{0.1cm}
    {\large Sparsification of Polynomials via Shifts ($\PolySparse_R$):}
    \vspace{0.1cm}\\
    \hline
    \vspace{0.1cm}
    Given a polynomial $f\in R[x_1, \ldots, x_n]$,  decide if there exists a vector $(a_1, \ldots, a_n)\in R^n$ such that $f(x_1+a_1, \ldots, x_n+a_n)$ has strictly fewer monomials with non-zero coefficients than $f(x_1, \ldots, x_n)$, or output ``FAIL'' if no such vector exists.\\
    \hline
  \end{tabular}
\end{center}

In this paper we show that the problem $\PolySparse_R$ is at least as hard as as checking if a given system of polynomial equations over $R[x_1,\ldots, x_n]$ has a solution (Hilbert's Nullstellensatz).

\noindent
\emph{Hilbert's Nullstellensatz:} Given a system $S$ of polynomial equations $S=\{f_1=0, \ldots, f_r=0\}$ over the polynomial ring $R[x_1, \ldots, x_n]$, we say that the system is satisfiable if there exists an assignment $\veca\in R^n$ to the variables that simultaneously satisfies all  equations in $S$. This problem has a great significance in Algebraic Geometry and has other important applications in diverse areas. We state a slightly restricted version of Hilbert's Nullstellensatz problem that asks for a common solution in a specific domain (the general version asks for a solution in the algebraic closure). This definition is similar to the definition in the Blum, Shub and Smale model of computation \cite{BSS98}.

\begin{center}
  \begin{tabular}{|m{15cm}|}
    \hline
    \vspace{0.1cm}
    {\large Hilbert's Nullstellensatz over a ring $R$ ($\HN_R$):}
    \vspace{0.1cm}\\
    \hline
    \vspace{0.1cm}
    Given a system of polynomial equations $\{f_1=0, \ldots, f_r=0\}$ over $R[x_1, \ldots, x_n]$, decide whether there exist a vector $(a_1, \ldots, a_n)\in R^n$ such that for all $i\in[r]$, $f_i(a_1, \ldots, a_n) = 0$, or output ``FAIL'' if no such vector exists.\\
    \hline
  \end{tabular}
\end{center}

With this background, we shall now state our  first main result that gives a reduction from Hilbert's Nullstellensatz problem to polynomial sparsification.
\begin{theorem}\label{thm:HN-hard}
  Let $R$ be an integral domain,  which is not a field. Then $\PolySparse_R$ is $\HN_R$-hard, in any of the white-box representations.
\end{theorem}

If $R$ is arbitrary, then the polynomials could have coefficients with arbitrary bit complexity. Thus, it is important for us to also specify the model of computation over which this problem is being considered. In the Turing machine model, assuming that $f_1,\ldots,f_r\in \C[x_1,\ldots, x_n]$ have integral coefficients, Koiran~\cite{Koi96} showed (by assuming that the Generalized Riemann Hypothesis is true) that $\HN_\mathbb{C}$ can be solved in the second level of polynomial hierarchy. Without the GRH assumption, the only known upper bound for $\HN_\mathbb{C}$ is  $\mathsf{PSPACE}$. 
For $R=\mathbb{Z}$, Matiyasevich~\cite{Mat70} showed that this problem is undecidable (also see \cite{Dav73}). Putting these together with \cref{thm:HN-hard} we get the following  consequence. 

\begin{corollary}\label{cor:HN-Sparse}
$\PolySparse_\mathbb{Z}$ is undecidable.
\end{corollary}

It is important to note that under sparse or dense representations, $\PolySparse_{R}$ is in $\NP_{{R}}$. That is, given $\veca$, we can efficiently verify if it is a sparsifying shift for a polynomial $P(X)$ using at most polynomially many algebraic operations using sparse polynomial interpolation, and sparse polynomial identity testing\footnote{Given a polynomial $P(X)$ (of sparsity $t$) in its sparse representation, it is easy to see that we have access to evaluations of the polynomial $P(X+\veca)$ as well. Let $Q(X)$ be the $(t-1)$-sparse polynomial obtained by polynomial interpolation (using \cite{BT88} which uses at most polynomially many algebraic operations) using these evaluations. If $\veca$ were a sparsifying shift, then sparsity of $P(X+\veca)$ would strictly be smaller than $t$ and in that case, $Q(X)$ would in fact be equal to $P(X+\veca)$. This equivalence can be checked using polynomial identity testing of sparse polynomials (using \cite{KS01} which also uses at most polynomially many algebraic operations).}. Thus for any integral domain $R$ (which is not a field) such that $\HN_R$ is $\NP_R$-complete, over the Blum-Shub-Smale model of computation~\cite{BSS98}, we get the following corollary from the aforementioned statements and \cref{thm:HN-hard}.

\begin{corollary}\label{cor:bss-npc}
 Let $R$ be an integral domain (but not a field) such that $\HN_R$ is $\NP_R$-complete over the Blum-Shub-Smale model of computation. Then $\PolySparse_{R}$ is also $\NP_{{R}}$-complete.
\end{corollary}
In particular, we get that $\PolySparse_{\Z}$ is also $\NP_{{\Z}}$-complete.\\
 
These results, to some extent, shed a light on why this problem in general has been evading the efforts to provide efficient algorithms.

Note that our problem $\PolySparse_R$ can  also be viewed as a gap decision problem -- given a polynomial $P(X)$ of sparsity $t$, is there a vector $\veca$ such that $P(X+\veca)$ has at most $t-1$ monomials. Let us formally define a more general gap version of $\PolySparse_R$.

\begin{center}
  \begin{tabular}{|m{15cm}|}
    \hline
    \vspace{0.1cm}
    {\large $\alpha$-gap-$\PolySparse_R$:}\vspace{0.1cm}\\
    \hline
    \vspace{0.1cm}
    Let $\alpha>1$ be a parameter. Given a polynomial $P\in R[X]$ and a parameter $t$,
    \begin{itemize}
    \item output YES if there exists a vector $\veca$ such that $P(X+\veca)$ has at most $t$ monomials, and
    \item output NO if for all vectors $\veca$, $P(X+\veca)$ has at least $\alpha t$ monomials.
    \end{itemize}\\
    \hline
  \end{tabular}
\end{center}
Using  gap amplification we reduce $\PolySparse_R$ to $N^\beta$-gap-$\PolySparse_R$ for all functions $\beta \in o(1)$. We thus get our second main result.

\begin{theorem}\label{thm:gap-hardnessZ}
For every function $\beta:\N\to\R_+$ such that $\beta\in o(1)$, $N^\beta$-gap-$\PolySparse_\mathbb{Z}$ is undecidable (where $N$ is the input length).

\end{theorem}

In \cref{thm:gap-hardnessZ}, we used the undecidability of $\HN_\Z$ to infer the undecidability of $N^\beta$-gap-$\PolySparse_\mathbb{\Z}$. However, we do not have such  results for  rings  $R\neq \mathbb{Z}$ (over Turing machine model). Furthermore, $HN_\Q$ is not known to be undecidable and, as mentioned above, over $\C$ it is decidable as well as over finite fields. Thus for $R=\F_p, \mathbb{Q}, \mathbb{R}$ or $\mathbb{Z}_q$, we present a different reduction of gap problems -- from $(1-\varepsilon, \delta)$-gap-$\maxthreelin_R$ to $\alpha$-gap-$\PolySparse_R$ and infer  $\NP$-hardness results for $\alpha$-gap-$\PolySparse_\mathbb{R}$.

\begin{center}
  \begin{tabular}{|m{15cm}|}
    \hline
    \vspace{0.1cm}
    {\large $(1-\varepsilon,\delta)$-gap-$\maxthreelin_R$:}\vspace{0.1cm} \\
    \hline
    \vspace{0.1cm}
    Given a system of linear equations $\{L_1=0, \ldots, L_m=0\}$ over $R[x_1, \ldots, x_n]$ each of which depends on exactly $3$ variables,
    \begin{itemize}
    \item output YES if at least $(1-\varepsilon)$ fraction of equations can be simultaneously satisfied, and
    \item output NO if at most $\delta$ fraction of equations can be simultaneously satisfied.
    \end{itemize}\\
    \hline
  \end{tabular}
\end{center}

We say that it is $\NP$-hard to $(1-\varepsilon, \delta)$-approximate $\maxthreelin_R$ 
if the decision problem $(1-\varepsilon,\delta)$-gap-$\maxthreelin_R$ is $\NP$-hard. Using this notion, we summarize non-exhaustively some known $\NP$-hardness results for $(1-\varepsilon, \delta)$-approximating $\maxthreelin_R$.

\begin{center}
  \begin{tabular}{|m{4.5cm}|m{3cm}|m{7cm}|}
    \hline
    \vspace{0.1cm}
    Result & Ring $R$ & $\NP$-Hardness for\\
    \hline
    H{\aa}stad~\cite{has01} & $\F_p$ & $\forall~\varepsilon>0$, $(1-\varepsilon, \frac{1+\varepsilon}{p})$-approximation\\
    \hline
    H{\aa}stad~\cite{has01} & $\mathbb{Z}_q$ for $q\in\mathbb{N}$ & $\forall~\varepsilon, \delta>0$, $(1-\varepsilon, \frac{1}{q}+\delta)$-approximation\\
    \hline
    Feldman, Gopalan, Khot and Ponnuswami~\cite{FGKP06} & $\mathbb{Q}$ & $\forall~\varepsilon>0$, $(1-\varepsilon, \varepsilon)$-approximation\\
    \hline
    Gurswami and Raghavendra~\cite{GR06, GR07} & $\mathbb{Q}, \mathbb{R}$& $\forall~\varepsilon, \delta>0$, $(1-\varepsilon, \delta)$-approximation\\
    \hline
  \end{tabular}
  \captionof{table}{Non-exhaustive list of known $\NP$-hardness results for approximating $\maxthreelin_R$}
  \label{table:max3lin-hardness}
\end{center}

Thus, a gap reduction from $(1-\varepsilon',\delta')$-gap-$\maxthreelin_R$ (where $\varepsilon'$ and $\delta'$ are as in the last column of \cref{table:max3lin-hardness}) to $\alpha$-gap-$\PolySparse_R$ (for $\alpha = \alpha(\varepsilon', \delta', R)$) implies  hardness of $\alpha$-gap-$\PolySparse_R$ and using amplification we get our third main result. We first state it in the sparse representation model and then in the arithmetic circuit model.

\begin{theorem}[Sparse representation]\label{thm:gap-sparse-input}
    For $R=\F_p, \mathbb{Q}, \mathbb{R}$ or $\mathbb{Z}_q$ and for every $\beta>1$  the $\beta$-gap-$\PolySparse_R$ problem is NP-hard. Furthermore, there exists a constant $\alpha>1$ such that for every $d=O(1)$ the $\alpha^d$-gap-$\PolySparse_R$  problem is  NP-hard when given polynomials of degree at most $d$ as input. 
\end{theorem}

The theorem is stated for the sparse representation model but as the polynomials under consideration  have many non-zero terms it can also be stated without any modification in the dense representation model. We next state the theorem in the arithmetic circuit model.

\begin{theorem}[Arithmetic circuit representation]\label{thm:gap-circuit-input}
  There exists a constant $\alpha>1$ such that the following holds for $R=\F_p, \mathbb{Q}, \mathbb{R}$ or $\mathbb{Z}_q$. For every $d$  the $\alpha^d$-gap problem is NP-hard when given polynomials of degree at most $d$ as input. Furthermore, our hard instances have circuit size $(nd+1)$.
\end{theorem} 

Observe that  if we take e.g. $d=n^c$ in the theorem above then the input size is $N=n^{c+1}$ and the gap is $\exp(N^{1-1/c})$.

\section{Preliminaries}
We use $[n]$ to refer to the set $\inbrace{1,2,\ldots, n}$.  We use capital letters $A$ and $C$ to represent matrices, capital letters $U,S$ and $T$ to represent systems of equations, and capital letters $X,Y$ and $Z$ to represent sets of variables. We reserve letters $x,y$ and $z$ with, or without subscripts, to represent variables. We use bold letters $\veca, \vecb, \ldots$ to indicate vectors and non-bold letters (apart from $x,y$ and $z$) with, or without subscripts, $e, b_i, a_j, A_{i,j}, C_{k, \ell}, \ldots$ to indicate scalars.

Let $S$ be a system of polynomial equations $\{f_i=0\}_{i=1}^r$, where $\deg(f_i)=d_i$. We use $\mathrm{Vars}(f)$ to denote the variable support of the polynomial $f$, and for a system $S$ of polynomial equations we use $\mathrm{Vars}(S)$ to denote the union of $\mathrm{Vars}(f_i)$ for all equations $f_i=0$ in $S$. We denote with $L$ an upper bound on the bit-complexity of the coefficients of the polynomials in the system.\footnote{When the underlying ring is an abstract ring one has to define this complexity, but for the usual rings and fields such as $\Z,\Q,\R,\C,\F_q$ this is the natural definition. In the BSS model this complexity is called height and is indeed only defined for these natural domains \cite{BSS98}.} The total degree of $S$ is
$d=\sum_i d_i$. 

In this paper we shall consider two types of representations of polynomials (and hence of polynomial equations). The representation that is typically studied in the context of polynomial equations is the so called ``sparse
representation''. In this representation polynomials are given as a set of pairs consisting of exponent vectors together with the coefficients of the corresponding monomials. E.g. the polynomial $2x^2z-y\in\F[x,y,z,]$ is represented as $\{((2,0,1),2),(0,1,0),-1)\}$. This is called the sparse representation as  we do not charge for monomials whose coefficients are equal to $0$. In particular the size of the representation of a degree $d$ polynomial can be much smaller than ${n+d \choose d}$. For a system of polynomial equations $S=\{f_i=0\}_{i=1}^{r}$, the complexity of $S$, or its size, is defined to be the total bit size of the sparse representations of the polynomials $\{f_i\}_{i\in[r]}$. We note that this is always upper bounded by $\sum_{i=1}^{r} {n+d_i \choose d_i}\cdot L$.

The second type of representation that we consider is when the polynomials $f_i$ are given as the outputs of arithmetic circuits.\footnote{Arithmetic circuits are directed acyclic graphs whose leaf nodes are labeled by variables or constants from the underlying field, and every non-leaf node is labeled either by a $+$ or $\times$. The fan-in of multiplication gates is $2$ while the fan-in of addition gates is unbounded. Every node computes a polynomial by operating on its inputs with the operation given by its label. The computation flows from the leaves to the output node (see \cite{SY10, sapt} for more details).} In this paper we only consider the white-box version of this representation, i.e., when the computation graph of the circuit is explicitly given to the algorithm. In this case the complexity (or size) of the system $S$ is the total size of the input circuits times the maximal bit complexity of  coefficients in the circuits.

As we shall later see (\cref{lem:sparserep-to-equations} and \cref{lem:circuit-to-equations}), given a system of equations, either via arithmetic circuits or in the sparse representation model, one can easily construct an equivalent system $T$ of polynomial equations of degree $2$ and roughly of the same complexity, such that the system $S$ has a solution if and only if the system $T$ does. Hence, these two different representations have the same computational power. However, this reduction is not gap-preserving so we will have to give separate arguments for the gap problems.

\section{Reduction from $\HN_R$ to $\PolySparse_R$}\label{sec:HN-hard}

In this section, we shall first show that given a system $S$ of $r$ many polynomial equations over $R[X]$, we can algorithmically construct a system $T$ of polynomial equations over $R[X''$] such that $X\subseteq X''$; each polynomial in $T$ is of degree at most $2$; and if $\veca\in R^{\abs{X}}$ is a solution for the system $S$ then there exists an extension $\veca'$ of $\veca$ such that $\veca'$ is a solution for the system $T$. And vice versa, from a solution to $T$ we deduce a solution to $S$. From $T$ we shall then construct a polynomial $P_S\in R[X'',W]$ such that $S$ has a solution if and only if the polynomial $P_S$ can be sparsified. 

\subsection{Reduction to a system of polynomial equations of degree at most $2$}\label{subsec:reduction-degree2}

\subsubsection{Case when input is provided in sparse representation}
Let  $S = \{f_1=0, \ldots, f_r=0\}$ be our input system of polynomial equations such that each $f_i$ is provided in sparse representation. 
Let $s_i$ denote the number of monomials with non-zero coefficients in the polynomial $f_i$. 
Let $Y$ and $Z$ be new disjoint sets of variables, that are disjoint from $X$ such that
\begin{align*}
   Y = \{y^{(i)}_{j,k}\mid i\in [r], j\in [s_i]~\text{and}~ k\geq 1 \}~\text{and}~ Z = \{z^{(i)}_{j}\mid i\in [r], j\in [s_i]\}.
\end{align*}

Let the variables in $Y$ have a lexicographic ordering based on the indices $i,j$ and $k$, variables in $Z$ have a lexicographic ordering based on the indices $i$ and $j$, and the variables in $X$ have some arbitrary ordering. Across the sets $X,Y$ and $Z$, let the ordering be $Z\succ Y \succ X$. Given $S = \{f_1=0, \ldots, f_r=0\}$, we construct an extended system $T$ of polynomial equations over the variables $X\sqcup Y\sqcup Z$ such that each polynomial equation is of degree at most $2$ and is such that there is a solution $\veca\in R^{\abs{X}}$  for the system $S$ if and only if there exists an extension $\veca'$ of $\veca$ such that $\veca'$ is a solution for the system $T$. This is a well known reduction (see, e.g., Lemma 6 in Chapter 2 of \cite{BSS98}) but for completeness we repeat it here.

\cref{alg:extendSystem} and \cref{alg:reduceMonomial} describe the construction of the extended system of equations. What the algorithms do is, roughly, for any monomial $m=x_{i_1}\cdot x_{i_2} \cdot x_{i_3}\cdot\ldots\cdot x_{i_j}$ of degree greater than $2$,  introduce a new variable, say $y$, replace $m$ with the the monomial $y\cdot x_{i_3}\cdot\ldots\cdot x_{i_j}$ and introduce a new equation $y-x_{i_1}\cdot x_{i_2}=0$ and for any monomial $m'$, of degree $1$, introduce a new variable $z$ and  a new equation $z-m' = 0$. Finally, an affine linear equation of the form $\sum c_i z_i+c_0$ is added to account for the fact that the original sum of monomials has to be zero. In particular, at the termination of the algorithm, the system $T$ consists of  constant-free quadratic binomial\footnote{We use the phrase constant-free quadratic binomial equation to refer to an equation with two non-constant monomials of degree  at most $2$.} equations and affine linear polynomial equations. It is also clear that there exists $\veca'$ that satisfies $T$ if and only if there is  $\veca$ that satisfies $S$. 

\begin{algorithm}[H]
  \caption{ConstructExtendedSystem-SparseRepresentation($S$)}\label{alg:extendSystem}
  \SetAlgoLined
  \KwResult{Given a system $S= \{f_1=0, \ldots, f_r=0\}$ of polynomial equations provided in sparse representation, we construct a system $T$ of polynomial equations over an extended set of variables such that each polynomial in $T$ has degree at most 2 and if $\veca$ solves $S$ then there exists an extension $\veca'$ of $\veca$ such that $\veca'$ solves $T$ and vice-versa.}
  $T\gets \emptyset$\;
  
  \For{$i\in [r]$}{
    $f' \gets 0$\;
    Let $s_i$ be the sparsity of $f_i$\;
    \For{$j\in[s_i]$}{
      Let $c_{i,j}$ be the coefficient of $j$'th monomial $m_{i,j}$ in $f_i$\;
      \eIf{$\deg(m_{i,j})\geq 1$}{
	$U \leftarrow \text{ReduceMonomial($m_{i,j}$, $i$, $j$)}$\;
	$T\gets T\cup U$\;
	$f'\gets f'+ c_{i,j}\cdot z^{(i)}_{j}$\;
      }
      {
	$f'\gets f'+ c_{i,j}$\;
      }
    }
    $T\gets T\cup \{f'=0\}$\;\label{line:f'}
  }
  \Return $T$
\end{algorithm}

\begin{algorithm}[H]
  \caption{ReduceMonomial($m$, $i$, $j$)}\label{alg:reduceMonomial}
  \SetAlgoLined
  \KwResult{Given the $j$'th monomial (of degree at least $1$) of the $i$'th polynomial, $m_{i,j}$, generate a collection of polynomial equations $U$ from it.}
  $U\gets \emptyset$\;
  $k = 1$\;
  \While{$\deg(m) \geq 2$}{
    Let $u,v$ be the renaming of the two trailing variables under the ordering of $X$ and $Y$ variables as described above\;
    
    $U\gets U\cup \{y^{(i)}_{j, k}- u\cdot v = 0\}$\;
    $m\gets \frac{m}{u\cdot v}\cdot y^{(i)}_{j, k}$\;
    $k \gets k+1$\;
  }
  $U\gets U\cup \{z^{(i)}_{j}- m = 0\}$\;\label{line:linear-2}
  
  \Return $U$
\end{algorithm}

\newpage

\begin{lemma}\label{lem:sparserep-to-equations}
  Let $S = \{f_i(X) = 0\}_{i=1}^r$ be a system of polynomial equations over the polynomial ring $R[X]$ such that for each $i\in[r]$, $f_i(X)$ is a polynomial of degree $d_i$ and sparsity $s_i$, and the bit complexity of each coefficient is at most $L$. Then, \cref{alg:extendSystem} runs in time $ \poly(\abs{X}, r, \max_i\{d_i\}, \max_i\{s_i\},L)$ and returns a set $T$ of polynomial equations $\{g_j(X,Y,Z) = 0\}_{j=1}^t$ over the polynomial ring $R[X, Y, Z]$ such that 
  \begin{itemize}
  \item $t = \poly(\abs{X}, r, \max_i\{d_i\}, \max_i\{s_i\},L)$.
  
  \item $\abs{X\sqcup Y\sqcup Z} = \poly(\abs{X}, t, \max_i\{d_i\}, \max_i\{s_i\})$.

  \item The bit-complexity of the coefficients of the polynomial equations in $T$ is also at most $L$. 
  \item Each polynomial equation $g_j(X,Y,Z) = 0$ in $T$ is either a quadratic binomial polynomial equation or an affine linear polynomial equation.
  \item $S$ has a solution $\veca\in R^{\abs{X}}$ if and only $T$ has a solution $\veca'\in R^{\abs{X\sqcup Y \sqcup Z}}$.
  \end{itemize} 
\end{lemma}

The lemma is very easy to verify and so we only give a brief proof of the last claim.

\begin{proof}
Note that by the aforementioned ordering of variables, in all quadratic binomial equations included into the set $T$ that have the form $u-v\cdot w=0$, we have that $u$ is of the form $y^{(i)}_{j,k}$, and $v$ and $w$ could be of the form $y^{(i)}_{j,k'}$ or $x_{i'}$, and  the term $u$ is leading with respect to the terms $v$ and $w$. Further, all the quadratic equations in the set $T$ can be assumed to have some sort of a \emph{topological order}.  Thus the values of all the $Y$ variables that appear in the variable support can be inductively inferred by just setting the $X$ variables. That is, if we want to satisfy all such equations, then the value of the term $u$ can be inferred from the value of terms $v$ and $w$ for every invocation of $u$, $v$ and $w$. 

Further note that some of the linear polynomial equations in $T$ take the form $u'-v' =0$ (from \cref{line:linear-2} of \cref{alg:reduceMonomial}) where $u'$ is of the form $z^{(i)}_{j}$, and $v'$ could be of the form $y^{(i)}_{j,k}$ or $x_{i'}$. Similar to the case above, the value of the term $u'$ can be inferred from the value of the term $v'$ for every invocation of $u'$ and $v'$, and the value of $v'$ is already fixed as the values of all the $X$ and $Y$ variables that appear in the variable support were set in the aforementioned discussion. Observe that the rest of the linear polynomial equations in $T$ (from \cref{line:f'}  of \cref{alg:extendSystem}) correspond to the polynomial equations in $S$, and setting of $Z$ variables in the variable support of $T$, by the above procedure, satisfies all the linear polynomial equations in $T$. 

Given an assignment $\veca$ to $X$ let $\veca'$ be its unique extension to the variable set $X\sqcup Y\sqcup Z$ according to the process described above.
The argument above shows that $\veca$ is a solution to the system $S$ if and only if $\veca'$ is a solution to the system $T$.
\end{proof}

\subsubsection{Case when input is provided in white-box circuit form}

Let $S = \{f_i=0\}_{i=1}^r$ be our input system of polynomial equations such that each $f_i$ (for $i\in[r]$) is provided as an arithmetic circuit $\Phi_i$ of size $s_i$. Without loss of generality, for all $i\in [r]$, we can assume that every product gate in $\Phi_i$ has a fan-in of $2$.

Let $Y  = \inbrace{y^{(i)}_{j}\mid i\in[r]~\text{and}~j\in[s_i]}$ be a new set of variables disjoint from $X$.
For $i\in [r]$, let $g^{(i)}_1, \ldots, g^{(i)}_{s_i}$ be a topologically sorted enumeration of all nodes in circuit $\Phi_i$. For all $j\in[s_i]$, let node $g^{(i)}_{j}$ be labelled by the variable $y^{(i)}_{j}$. Corresponding to each node in $\Phi_i$, we shall now define a polynomial equation of degree at most $2$ over the variable sets $X$ and $Y$.

\begin{algorithm}[!t]
  \caption{ConstructExtendedSystem-CircuitRepresentation(S)}\label{alg:extendSystem-circuit}
  \SetAlgoLined
  \KwResult{Given a system $S= \{f_1=0, \ldots, f_r=0\}$ of polynomial equations provided in white-box circuit representation, we construct a system $T$ of polynomial equations over an extended set of variables such that each polynomial in $T$ has degree of at most 2 and if $\veca$ solves $S$ then there exists an extension $\veca'$ of $\veca$ such that $\veca'$ solves $T$ and vice-versa.}
  $T\gets \emptyset$\;
  
  \For{$i\in [r]$}{ 
    Let $\Phi_i$ be the circuit computing $f_i$, and $s_i = \abs{\Phi_i}$\;
    \For{$j\in[s_i]$}{ 
        \eIf{$g_j^{(i)}$ is a leaf node in $\Phi_i$}{
        Let $u$ be a variable or a constant labeling the input node in $\Phi_i$\;
        $T \gets T \cup \{y^{(i)}_j - u =0\}$\;
      }{
      \eIf{$g^{(i)}_{j}$ is a product node in $\Phi_i$}{
        Let $u$ and $v$ be the renaming of the labels of the children of $g^{(i)}_{j}$\;
         $T \gets T \cup \{y^{(i)}_j -  u\cdot v = 0\}$\;
      }{
        $g^{(i)}_{j}$ is a sum node in $\Phi_i$\;
        Let $u_1, \ldots, u_k$ be the renaming of the labels of the children of $g^{(i)}_{j}$\;
        $T \gets T \cup \{y^{(i)}_j - \sum_{i=1}^k  u_i= 0\}$\;
      }
    }
  }
  }
  \Return $T$\;
\end{algorithm}

\cref{alg:extendSystem-circuit} describes the construction of the extended system of equations. For each input node $g^{(i)}_{j}$, labeled by  $u$ (where  $u$ is either a variable $x_{k}$ or a constant $c$) in $\Phi_i$, the algorithm adds the polynomial equation  $y^{(i)}_j - u= 0$ to the system $T$. For each product node $g^{(i)}_{j}$ with children labelled $u$ and $v$ (where $u,v$ could be of the form $y^{(i)}_{j'}$ for some $j>j'$ or $x_{k'}$), the algorithm introduces a new polynomial equation $y^{(i)}_j - u\cdot v = 0$ to the system $T$, and for each sum node $g^{(i)}_{j}$ with children labelled $u_1, \ldots, u_k$ (where $u_1, \ldots, u_k$ could be of the form $y^{(i)}_{j'}$ for some $j>j'$ or $x_{k'}$), it introduces a new polynomial equation $y_j^{(i)} - \sum_{i=1}^k  u_i = 0$. At the termination of the algorithm, the system $T$ consists of  either constant-free quadratic binomial equations or affine linear polynomial equations. It is also clear that there exists $\veca'$ that satisfies $T$ if and only if there is  $\veca$ that satisfies $S$. As before this is easy to see: by following the flow of computation in an arithmetic circuit from leaves to the root, we can infer the values of $y^{(i)}_j$ for all $i\in[t]$ and $j\in [s_i]$.

\begin{lemma}\label{lem:circuit-to-equations}
  Let $S = \{f_i(X) = 0\}_{i=1}^r$ be a system of polynomial equations over the polynomial ring $R[X]$ such that for each $i\in[r]$, the polynomial $f_i(X)$ is provided as an arithmetic circuit $\Phi_i$ of size $s_i$. Then, when given this as input, \cref{alg:extendSystem-circuit} runs in time $\poly(\abs{X}, r, \max_i\{s_i\})$ and returns a system $T$ of polynomial equations $\{g_j(X,Y) = 0\}_{j=1}^t$ over the polynomial ring $R[X, Y]$ such that 
  \begin{itemize}

  \item $t = \poly(\abs{X}, r, \max_i\{s_i\})$.
    \item $\abs{Y} \leq r\cdot\max_i\{s_i\}$.
  \item Each polynomial equation $g_j (X,Y) = 0$ in $T$ is either a quadratic binomial polynomial equation or an affine linear polynomial equation.
  \item $T$ has a solution in $R^{\abs{X\sqcup Y}}$ if and only $S$ has a solution in $R^{\abs{X}}$.
  \end{itemize} 
\end{lemma}

\subsection{Construction of  $P_S$}\label{subsec:poly-defn}

Given a system $S$ of polynomial equations over a set of variables $X$, in \cref{subsec:reduction-degree2} we constructed the system $T$ of polynomial equations of degree at most $2$ over the set of variables $X, Y$ and $Z$,\footnote{In case the polynomials in the system $S$ of polynomial equations are provided as circuits, $Z = \emptyset$.} such that $\abs{X\sqcup Y\sqcup Z}$ is at most polynomial in the input size. Without loss of generality, let the variables in $X\sqcup Y \sqcup Z$ be renamed as the variable set $X' = \inbrace{x_1,\ldots, x_N}$ where $N = \abs{X\sqcup Y\sqcup Z}$. 

Let $\{g_1(X')=0, \ldots, g_t(X')=0\}$ be the enumeration of polynomial equations in $T$. 
Without loss of generality, we can assume that the number of equations with a non-zero constant term is equal to $1$. Otherwise, given a system $T$ of polynomial equations, with $t'>1$ many of these polynomial equations with non-zero constant terms, we shall construct a new system $T'$ such that the number of polynomial equations in $T'$ that have non-zero constant terms is exactly equal to $1$, and a solution of $T$ is a solution of $T'$ and vice versa. Without loss of generality assume that $\{g_1(X')=0, \ldots, g_{t'}(X') =0\}$ are the polynomial equations in $T$ with non-zero constant terms. By Lemmata~\ref{lem:sparserep-to-equations} and~\ref{lem:circuit-to-equations} it follows that $\{g_1(X')=0,\ldots,g_{t'}(X')=0\}$ are affine linear equations.
Denote the free term in $g_1(X'),\ldots,g_{t'}(X')$ with  $c_1, \ldots, c_{t'}$, respectively. We obtain $T'$ from $T$ by just updating each of the polynomials $g_i(X')$ (for $2\leq i\leq {t'}$) as follows. 
\begin{align*}
    g_i(X') \gets c_1\cdot g_i(X') - c_i\cdot g_1(X')\,.
\end{align*}
The rest of the polynomials from $T$ are directly added to $T'$. It is easy to see that any solution to the system $T$ is also a solution to the system $T'$ and vice-versa (as $R$ is a domain). Furthermore, the only equation in $T'$ with a non-zero constant term is an affine linear equation.

Conditioned on the aforementioned discussion, we shall assume that all polynomial equations in $T$ other than $g_1(X') = 0$, have no constant terms. For a new variable $x_0$, let $X'' = X' \sqcup \{x_0\}$ and thus $\abs{X''} = N+1$. Let $W=\inbrace{w_1,\ldots, w_{t}}$ be a new set of variables disjoint from $X''$. Let $\gamma$ be an element in $R$ without a multiplicative inverse (recall that in \Cref{thm:HN-hard} we assume that $R$ is not a field). We shall now define our polynomial $P_S$ in the polynomial ring $R[X'',W]$ as follows
\begin{equation}\label{eqn:poly-defn}
  P_S(X'', W) = \underbrace{ w_1\cdot g_1(X')}_{I} + \underbrace{\inparen{\sum_{i =2}^t w_i\cdot\inparen{ \gamma\cdot g_i(X') + \sum_{k=0}^Nx_k}}}_{II}\,. 
\end{equation}
Observe that $\deg(P_s)\leq 3$.

\begin{remark}\label{rem:poly-complexity}
  The sparsity of the polynomial $P_S(X'',W)$, $\sigma$, is equal to the sum of sparsities of polynomials in each of its summands, and it is equal to $(t-1)\cdot (N+1)+ \sum_{i=1}^t s'_i $ where $s'_i$ is the sparsity of the polynomial $g_i(X')$. On the other hand, $P_S(X'',W)$ can also be represented as a depth four arithmetic circuit, with at most $3t$ non-leaf nodes.
\end{remark}

We shall now show that it is sufficient to consider shifts with a certain structure for $P_S$. Further we shall show that a solution to the system $S$ of polynomial equations exists if and only if there exists a vector $\vecb$ such that $P_S(X''+\vecb,W)$ has fewer monomials than $P_S(X'', W)$.

\begin{lemma}
  \label{lem:shift-structure}
  Let $\veca = \inbrace{a_1, \ldots, a_N} \in R^{N}$ be a solution to the system $T$ of polynomial equations. Let $\vecb = \inbrace{b_0, \ldots, b_N}, \vecb' =\inbrace{b'_0, \ldots, b'_N} \in R^{N+1}$ and $\veca' =\inbrace{a'_0, \ldots, a'_N}$ be such that
  \begin{itemize}
  \item $b_i = b_i'$ for all $i\in [N]$,
  \item $b'_0 \neq b_0$ and $b_0 = - \sum_{i\in [N]} b_i$,
  \item $a_i' = a_i$ for all $i\in [N]$, and
  \item $a'_0 = - \sum_{i\in [N]} a_i$.
  \end{itemize}
  Let $\vecb''$ and $\vecc$ be any vectors in $R^{N+1}$ and $R^{t}$ respectively. Then,
  \begin{enumerate}
  \item The sparsity of $P_S(X''+\vecb'', W+\vecc)$ is at least that of $P_S(X''+\vecb'', W)$, \label{item:1}
  \item The sparsity of $P_S(X''+\vecb', W)$ is at least that of $P_S(X''+\vecb, W)$.
  \label{item:2}
  \item The sparsity of $P_S(X'', W)$ is $1$ more than that of $P_S(X''+\veca',W)$.\label{item:3}
  \end{enumerate}
\end{lemma}

\begin{proof}
  Given the structure of the polynomial $P_S(X'',W)$, proof of \Cref{item:1} follows directly from the fact that the polynomial $P_S(X'',W)$ is linear in the $W$ variables and thus all terms of $P_S(X''+\vecb'', W)$ also appear in $P_S(X''+\vecb'', W+\vecc)$. Further, the difference  $P_S(X''+\vecb'', W+\vecc) - P_S(X''+\vecb'', W)$ does not depend on any $W$ variable.

  From their definition, the vectors $\vecb$ and $\vecb'$ are identical when projected down to their last $N$ coordinates and these exactly correspond to shifts of variables in $X'$. We shall use $\vecb|_{X'}$ to denote this projection. In particular, for all $i\in[t]$,
  $g_i(X'+\vecb|_{X'}) = g_i(X'+\vecb'|_{X'})$.
  It is easy to see that the polynomial $\sum_{i=0}^N x_i$ is invariant under shift by $\vecb$ (as $\sum_{i=0}^N b_i = 0$) but not under shift by $\vecb'$. Putting both of these facts together we can now say that $P_S(X''+\vecb', W)$ contains all the terms that are contained in $P_S(X''+\vecb, W)$, and it additionally contains a non-trivial linear polynomial in the $W$ variables. This proves \Cref{item:2} of the lemma.

  Towards proving \Cref{item:3} of the lemma, we claim that under a shift by $\veca'$, as defined in the statement of the lemma, sparsity of part $II$ in \cref{eqn:poly-defn} does not change, and sparsity of part $I$ definitely decreases. 

  All polynomial equations $\{g_i(X') = 0\mid 2 \leq i\leq t\}$, are constant free and can either be quadratic binomial polynomial equations of the form $(x_{p} -  x_q\cdot x_e) = 0$ (for some $p,q,e\in [N]$) or homogeneous linear polynomial equations of the form $\sum_{j=1}^k c_{i_j} x_{i_j} = 0$ (for some $i_1, \ldots, i_k\in [N]$ and scalars $c_{i_j}$).

  \emph{When the equation is a quadratic binomial polynomial equation:} Since $\veca = \veca'|_{X'}$ solves $T$ we get that $a'_p- a'_q\cdot a'_e = 0 $ and using this fact we can show that for each summand of this kind in part II, the sparsity does not change.
  \begin{align*}
    &\gamma\cdot ((x_{p}+ a'_p) - (x_q+a'_q)\cdot (x_e+a'_e)) + \sum_{i=0}^N(x_{i}+a'_i)\\ 
    &= \gamma\cdot ((x_{p} -  x_q\cdot x_e) -  (a'_q x_e +a'_e x_q)) + \sum_{i=0}^N x_{i} \qquad \text{(Since $a'_p- a'_q\cdot a'_e = 0 $ and $\sum_{i=0}^N a'_i = 0$)}\\
    &= \gamma\cdot(x_{p} -   x_q\cdot x_e) + \inparen{\sum_{i\in [N]\setminus \{q,e\}}x_{i}} + (1-\gamma\cdot   a'_q)\cdot x_e+ (1-\gamma\cdot    a'_e)\cdot x_q.
  \end{align*}
  Since $\gamma$ has no multiplicative inverse, neither $(1-\gamma\cdot    a'_q)$ nor $(1-\gamma\cdot   a'_e)$ can be equal to $0$.
  
  \emph{When the polynomial equation is a homogeneous linear polynomial equation:}  Since $\veca = \veca'|_{X'}$ solves $T$  we get that $\sum_{j=1}^k c_{i_j}a'_{i_j} = 0$ and thus the sparsity remains invariant for such summands in part $II$.

Finally consider the non-homogeneous linear polynomial equation $g_1(X')=0$. Without loss of generality, let $g_1(X') = c_1 x_1+\ldots+ c_kx_k+c$. Note that $\sum_{i=1}^k c_i a'_i+c =0$ as $\veca = \veca'|_{X'}$ solves $T$ and thus sparsity reduces by $1$ under shift by such a vector $\veca'$:
$$\sum_{i=1}^k c_i (x_i+a'_i)+c=\sum_i c_i x_i+\left(\sum_{i=1}^k c_ia'_i+c\right)=\sum_{i=1}^k c_i x_i \;.$$

  By putting together the analysis for all the summands we get that the polynomial $P_S(X''+\veca', W)$ has one monomial less than $P_S(X'',W)$. 
\end{proof}

\begin{lemma}
  \label{lem:shift-implies-solution}
  Let $\vecb = (b_0, b_1, \ldots, b_N) \in R^{N+1}$ such that $b_0 = -\sum_{i=1}^N b_i$, be a shift that sparsifies the polynomial $P_S(X'',W)$ by at least one monomial. Let $\veca\in R^N$ be the projection of vector $\vecb$ to its last $N$ coordinates. Then $\veca$ solves $T$.
\end{lemma}
\begin{proof}

   The proof of \cref{lem:shift-structure} shows that given the structure of the shift $\vecb$, a reduction in sparsity can only come from $g_1(X')$. That is, the sparsity of polynomials $g_i(X')$ for $i\geq 2$, can only increase upon a shift.
  
  For the sake of contradiction, let us assume that there exists a polynomial equation in $T$ that is not satisfied by $\veca$. 
  If for some $i\geq 2$, $g_i(X') = 0$ is a polynomial equation that is not satisfied by $\veca$, then this contributes an increase of $1$ to sparsity of the polynomial $P_S(X'',W)$ upon the shift by $\vecb$ (by adding a term of the form $c\cdot w_i$). Else if $g_1(X') =0$ is not satisfied by $\veca$, then there is no contribution to reduction in sparsity from part $I$. This is due to the fact that the term of the form $c\cdot w_1$ vanishes upon a shift by $\vecb$ if and only if $\veca$ solves $g_1(X') =0$. In either of these cases, the sparsity of $P_s(X''+\vecb, W)$ is not strictly less than that of $P_S(X'',W)$. This contradicts our assumption that $\vecb$ sparsifies $P_S(X'',W)$ by at least one monomial.
\end{proof}

By putting together Lemmata \ref{lem:sparserep-to-equations}, \ref{lem:circuit-to-equations}, \ref{lem:shift-structure}, and \ref{lem:shift-implies-solution}, we get the following formal statement.

\begin{theorem}[$\HN_R$ reduces to $\PolySparse_R$]\label{thm:reduction}
  Given a system $S$ of polynomial equations over the polynomial ring $R[X]$, there exists a polynomial $P_S(X'', W)\in R[X'', W]$ (where $X\subseteq X''$) such that the system $S$ is solvable if and only if there exists a shift that sparsifies the polynomial $P_S$ by a monomial. Furthermore, the size of the polynomial instance $P_S$ is polynomially related to the input size of the system $S$ of polynomial equations. This holds true in both the sparse-representation and circuit-representation.
\end{theorem}

We thus get that if $\PolySparse_R$ can be solved efficiently (in general) then $\HN_R$ can also be solved efficiently. In other words, $\PolySparse_R$ is at least as hard as $\HN_R$. This completes the proof of \cref{thm:HN-hard}. Putting \cref{thm:HN-hard} together with the fact that $\HN_\mathbb{Z}$ is undecidable (due to \cite{Mat70}).

\section{Undecidability of $\beta$-gap-$\PolySparse_\mathbb{Z}$ problem}\label{sec:approx-hard-Z}

Note that $\PolySparse_R$ can be rephrased as the following gap problem -- given a polynomial of sparsity $\sigma$, decide if there is a shift that sparsifies the polynomial to at most $\sigma-1$ monomials, or there is no shift that sparsifies the polynomial below $\sigma$ monomials. We shall now show a reduction from this ($\PolySparse_R$ problem) to $\beta$-gap-$\PolySparse_R$ for any $\beta>1$.

Let the sets $X''$ and $W$ be as defined in the construction of the polynomial $P_S$ in \Cref{subsec:poly-defn}. Let $d$ be a parameter that we shall soon fix. Let $X^{(1)}, \ldots, X^{(d)}$ and $W^{(1)}, \ldots, W^{(d)}$ be $d$ many disjoint copies of variable sets $X''$ and $W$ respectively. Let $X_d = \sqcup_{k=1}^d X^{(i)}$ and $W_d = \sqcup_{k=1}^d W^{(i)}$. For the sake of brevity, let us use the following notation: Let $Y = X''\sqcup W$. For all $k\in[d]$, let $Y^{(k)} = X^{(k)}\sqcup W^{(k)}$ and $\abs{Y^{(k)}} = N'$. Let $Y_d = \sqcup_{k=1}^d Y^{(i)}$, so that $|Y_d|=N'd$.
Let the polynomial $Q_d(Y_d)$ be defined as follows.
\begin{align}\label{eqn:amplification}
    Q_d(Y_d) = \prod_{k=1}^d P_S(Y^{(k)}).
\end{align}
Observe that the sparsity of $Q_d(Y_d)$ is given by the product of sparsities of $d$ many instances of $P_S(X'',W)$. 

\begin{lemma} \label{lem:gap-amplification}
  Let $P_S(X'',W)$ and $Q_d(Y_d)$ be the polynomials as defined above. Let $\sigma$ be equal to the sparsity of the polynomial $P_S$. Then,
  \begin{enumerate}
        \item $\deg(Q_d)\leq 3d$.
      \item $Q_d(Y_d)$ has sparsity equal to $\sigma^d$.
      \item If $P_S$ has a depth four circuit of size $s\leq 3t+N'$ (recall Remark~\ref{rem:poly-complexity}) then $Q_d$ has a depth five circuit of size $sd+1$.
      \item There is a vector $\veca\in R^{N'}$ such that $P_S(Y+\veca)$ has at most $\sigma-1$ monomials if and only if there exists a vector $\veca_d \in R^{N'\cdot d}$ such that $Q_d(Y_d+\veca_d)$ has at most $(\sigma-1)^d$ monomials.
      \item For all vectors $\veca\in R^{N'}$, $P_S(Y+\veca)$ has at least $\sigma$ monomials if and only if for all vectors $\veca_d \in R^{N'\cdot d}$  $Q_d(Y_d+\veca_d)$ has at least $\sigma^d$ monomials.
  \end{enumerate}
\end{lemma}
\begin{proof}
  The claim regarding the degree of $Q_d$ follows immediately from the fact that $\deg(P_S)\leq 3$.
  Given that $P_S(X'',W)$ has a sparsity of $\sigma$ and since $Q_d(Y_d)$ is defined to be a product of $d$ distinct copies of $P_S(X'',W)$, sparsity of $Q_d(Y_d)$ is equal to $\sigma^d$. Similarly, if $P_S$ can be computed by  a circuit of size $s$, then there is a depth five circuit of size $(sd+1)$ that computes the polynomial $Q_d(Y_d)$ -- its output node is a product node into which $d$ copies of circuits of $P_S(X'',W)$ feed into.
  
  If there is a vector $\veca\in R^{N'}$ such that $P_S(Y+\veca)$ has at most $\sigma-1$ monomials then by taking $\veca_d$ to be the concatenation of $\veca$, $d$ many times, we get that $Q_d(Y_d+\veca_d)$ has at most $(\sigma-1)^d$ monomials. If there is $\veca_d \in R^{N'\cdot d}$ such that $Q_d(Y_d+\veca_d)$ has at most $(\sigma-1)^d$ monomials then it cannot happen that there is no $\veca\in R^{N'}$ such that $P_S(Y+\veca)$ has at most $\sigma-1$ monomials. 
  
  If for all vectors $\veca\in R^{N'}$, $P_S(Y+\veca)$ has at least $\sigma$ monomials, then $Q_d(Y_d+\veca_d)$ must have at least $\sigma^d$ monomials for all $\veca_d\in R^{N'\cdot d}$.
  On the other hand if for all vectors $\veca_d \in R^{N'\cdot d}$, $Q_d(Y_d+\veca_d)$ has at least $\sigma^d$ monomials,  (for the sake of contradiction) let us suppose that there is a vector $\veca'\in R^{N'}$ such that $P_S(Y+\veca')$ has at most $\sigma-1$ monomials. As before (due to the product structure of $Q_d$) we get that there is a corresponding vector $\veca'_d\in R^{N'\cdot d}$ such that $Q_d(Y_d+\veca'_d)$ has at most $ (\sigma-1)^d$ monomials. This contradicts our assumption. Thus, for all $\veca\in R^{N'}$, $P_S(Y+\veca)$ has at least $\sigma$ monomials.
\end{proof}

\begin{theorem}\label{thm:HN-to-gapZ}
    Let $R$ be an integral domain but not a field.
  Given a system $S$ of polynomial equations over the polynomial ring in $n$ variables $R[X]$, for any function $\beta:\N\to\R_+$ such that $\beta\in o(1)$, there exist $d = d(S, \beta)\in \mathbb{Z}_{>0}$ and a polynomial $Q_d(Y_d)$, in $N'd$ variables, of degree at most $3d$, such that the system $S$ is solvable if and only if the $M^\beta$-gap-$\PolySparse_R$ problem for $Q_d(Y_d)$ is solvable, where $M$ is the representation length of $Q_d(Y_d)$ in the sparse representation.\footnote{Recall that in sparse representation polynomials are given as a set of pairs consisting of exponent vectors together with the coefficient of the corresponding monomial. Thus, $M\approx (\#\text{ monomials in } Q_d) \times ((N'd)\times O(\log d)) \times O(d \cdot b)$, where $b$ is the maximal bit complexity of a coefficient in $P_S(X'', W)$.} 
\end{theorem}

\begin{proof}
  Given a system $S$ of polynomial equations, we can construct the polynomial $P_S(X'',W)$ (as defined in \cref{eqn:poly-defn}). Let $\sigma$ be the sparsity of $P_S(X'', W)$. Recall from \cref{thm:reduction} that system $S$ has a solution if and only if there exists a shift that sparsifies the polynomial $P_S(X'',W)$ by a monomial. 
  
    Recall that $Q_d(Y_d)$ has $M'=\sigma^d$ many monomials. Let $\alpha = \frac{\sigma}{\sigma-1}$. 
    By putting together \cref{thm:reduction} and \cref{lem:gap-amplification}, we get that the system $S$ of polynomial equations is solvable if and only if the $\alpha^d$-gap problem for $Q_d(Y_d)$ is solvable. Calculating we get that $\alpha^d= \left(\frac{\sigma}{\sigma-1}\right)^d \approx e^{d/(\sigma-1)}=M'^{\frac{1}{(\sigma-1)\log \sigma}}$. Picking $d$ large enough so that $M'\geq\sqrt{M}$ and  $\beta(M)<\frac{1}{2(\sigma-1)\log \sigma}$ the claim follows.
  \end{proof}  

Putting \cref{thm:HN-to-gapZ} together with the fact that $\HN_\mathbb{Z}$ is undecidable (due to \cite{Mat70}) we get \cref{thm:gap-hardnessZ}. 

\section{Hardness of $\beta$-gap-$\PolySparse_R$ problem for $R=\F_p, \mathbb{Q}, \mathbb{R}$, or $\mathbb{Z}_q$}\label{sec:approx-hard-K}

In this section we prove Theorems~\ref{thm:gap-sparse-input} and~\ref{thm:gap-circuit-input} by giving a reduction from  $\maxthreelin_R$  to the $\alpha$-gap-$\PolySparse_R$ problem, for different domains $R=\F_p, \mathbb{Q}, \mathbb{R}$ or $\mathbb{Z}_q$. Observe that we now do not require that our ring $R$ is not a field. 

Let $X = \{x_1, \ldots, x_n\}$. Let the given system $S$ of linear equations be $\{L_1(X)=0, \ldots, L_m(X)=0\}$, where $L_i(X)\in R[X]$, and each equation $L_i(X)=0$ depends on exactly $3$ variables. Let the given system of equations be expressed together as $A\cdot X + \vecb = \mathbf{0}$ such that for all $i\in[m]$, $A_i\cdot X + b_i = 0$ is the $i$'th linear equation $L_i(X)=0$, where $A_i$ is the $i$'th row of the matrix $A$. Note that there are exactly three non-zero entries in each row of $A$. Let $w = \max\{2n, 2m\}$. Let $C$ be a $w\times w$ matrix such that
\begin{align*}
  \text{for all $1\leq i,j \leq w$}, \quad C_{i,j} =
  \begin{cases}
    A_{i,j-w+n} & \text{if $i \leq m$ and $j\geq w-n+1$;}\\
    0 & \text{otherwise.}
  \end{cases}
\end{align*}
In other words, $A$ is the top right block of $C$ and the rest of $C$ is zeros. 
Let $\vece = (e_1, \ldots, e_w)\in R^{w}$ be such that $e_i = b_i$ for all $1\leq i\leq m$ and $e_i=0$ otherwise. Let $e_0$ be some constant. Let $Y=\{y_1, \ldots, y_w\}$ be a new set of variables disjoint from $X$. Let the polynomial $Q_S(Y)\in R[Y]$ be defined as follows.
\begin{align*}
  Q_S(Y) = \sum_{i,j\in[w]}C_{i,j}\cdot y_iy_j + \sum_{i\in [w]}e_i \cdot y_i + e_0\,.
\end{align*}
Note that there are at most $3m$ many non-zero entries in $C$, and there are at most $m$ many non-constant linear terms. Thus the sparsity of this polynomial is at most $4m+1$. 

For some vector $\veca = (a_1, \ldots, a_w) \in R^{w}$ let us  examine the structure of the polynomial $Q_S(Y+\veca)$.

\begin{align*}
  &Q_S(Y+\veca)
  = \sum_{i,j\in[w]}C_{i,j}\cdot(y_i+a_i)(y_j+a_j) + \sum_{i\in [w]}e_i\cdot (y_i+a_i) + e_0\\
  &= \sum_{i,j\in[w]}C_{i,j}\cdot(y_iy_j +a_iy_j+a_jy_i+a_ia_j) + \sum_{i\in [w]}e_i\cdot (y_i+a_i) + e_0\\
  &= \sum_{i,j\in[w]}C_{i,j}\cdot(y_iy_j+a_ia_j) +\sum_{i,j\in[w]}a_i\cdot y_j\cdot C_{i,j}+ \sum_{i,j\in[w]}a_j\cdot y_i\cdot C_{i,j} + \sum_{i\in [w]}e_i\cdot (y_i+a_i) + e_0\\
  &= \sum_{i,j\in[w]}C_{i,j}\cdot y_iy_j + \sum_{i\in [w]}y_i\cdot(e_i + \sum_{j\in[w]}a_{j}\cdot(C_{i,j}+C_{j,i})) + \sum_{i,j\in[w]}C_{i,j}\cdot a_ia_j + \sum_{i\in[w]}a_i\cdot e_i + e_0\,.
\end{align*}

Observe that the quadratic part of the polynomial $Q_S(Y)$ remains unperturbed under the shift but the affine linear part of it could get perturbed. We shall now show that every non-zero coefficient in the linear part corresponds to a linear equation in $S$.
\begin{lemma}\label{lem:equivalence-approx}
  Let $\veca = (a_1, \ldots, a_w)\in R^w$. Let $\veca' \in R^n$ be the projection of $\veca$ down to its last $n$ elements, that is, for all $j\in [n]$, $a'_j = a_{j+w-n}$. Then, for all $i\in [m]$, the coefficient of $y_i$ in $Q_S(Y+\veca)$ is zero if and only if $\veca'$ satisfies the $i$'th linear equation $L_i(X)=0$. Moreover, for all $i>m$, the coefficient of $y_i$ is zero.
\end{lemma}
\begin{proof}
  For all $i\in [m]$, the coefficient of $y_i$ in the polynomial $Q_S(Y+\veca)$ is equal to $$e_i + \sum_{j\in[w]}a_{j}(C_{i,j}+C_{j,i})\,.$$ Note that for this regime of $i\in[m]$, $e_i = b_i$, $C_{j,i} = 0$, and $C_{i,j} = A_{i,j-w+n}$ for $j$ in $[w-n+1, w]$ and zero otherwise. Thus, the coefficient of such a $y_i$ reduces as follows.
  \begin{align*}
   \text{coef. of}~y_i~\text{in}~Q_S(Y+\veca) =  e_i + \sum_{j\in[w]}a_{j}(C_{i,j}+C_{j,i}) = b_i + \sum_{j'=1}^na_{w-n+j'}A_{i,j'} = b_i + \sum_{j'=1}^na'_{j'}A_{i,j'}.
  \end{align*}
  This is exactly the value obtained by evaluating the $i$'th linear polynomial $L_i$ at $\veca'$. Thus, we get that coefficient of $y_i$ (for $i\in[m]$) is zero if and only if $\veca'$ satisfies the $i$'th linear equation, i.e.,\ $L_i(\veca')=0$.

  For all $i>m$, $e_i=0$ and $C_{i,j} =0$. Further, $\veca_j = 0$ for all $j\leq w-n$. Hence,
  \begin{align*}
    e_i + \sum_{j\in[w]}a_{j}(C_{i,j}+C_{j,i}) = \sum_{j\in[w]}a_{j}\cdot C_{j,i} = \sum_{j=w-n+1}^w a_{j}\cdot C_{j,i} =0\,.
  \end{align*}
  The last equality in the math block above is due to the fact that the entries $C_{j,i}$ are equal to zero for $j\geq w-n+1$ and $i\geq m+1$ regardless of whether $n\geq m$ or $m\geq n$, from the construction of the matrix $C$. Thus the coefficients of the terms $y_i$ for all $i>m$ are zero.
\end{proof}

Using this correspondence, we can show the following reduction.
\begin{lemma}\label{lem:gap-reduction}
  Let $\veca = (a_1, \ldots, a_w)\in R^w$. Let $\veca' \in R^n$ be the projection of $\veca$ down to its last $n$ elements, that is, for all $j\in [n]$, $a'_j = a_{j+w-n}$. Then
  \begin{enumerate}
  \item $\veca'$ satisfies at most $\delta$ fraction of equations in $S$ if and only if $Q_S(Y+\veca)$ has at least $(4-\delta)m$ non-constant monomials, and
  \item $\veca'$ satisfies at least $1-\varepsilon$ fraction of equations in $S$ if and only if $Q_S(Y+\veca)$ has at most $(3+\varepsilon)m+1$ monomials.
  \end{enumerate}
\end{lemma}
\begin{proof}
  From the aforementioned discussion, the sparsity of the polynomial $Q_S(Y+\veca)$ is decided by the coefficients of the linear terms. Further, \cref{lem:equivalence-approx} characterizes that the coefficient of a linear term is zero if and only if the \emph{corresponding} linear polynomial equation is \emph{satisfied}. Thus at most $\delta$ fraction of equations in $S$ are satisfied if and only if at most $\delta$ fraction of coefficients of linear terms are equal to zero. In other words, if at least $(1-\delta)$ fraction of coefficients of linear terms are non-zero. The constant term $e_0$ could get cancelled out, that is, $\sum_{i,j\in[w]}C_{i,j}\cdot a_ia_j + \sum_{i\in[w]}a_i\cdot e_i+e_0$ could be zero.  
  Thus, $\veca'$ satisfies at most $\delta$ fraction of equations in $S$ if and only if $Q_S(Y+\veca)$ has at least $3m+(1-\delta)m = (4-\delta)m$ non-constant monomials.
  
  Similarly at least $1-\varepsilon$ fraction of equations in $S$ are satisfied if and only if at least $1-\varepsilon$ fraction of coefficients of linear terms are equal to zero. In other words, if at most $\varepsilon$ fraction of coefficients of linear terms are non-zero. Thus, $\veca'$ satisfies at least $1-\varepsilon$ fraction of equations in $S$ if and only if $Q_S(Y+\veca)$ has at most $(3+\varepsilon)m+1$ monomials.
\end{proof}

We first prove a more restricted version of Theorems~\ref{thm:gap-sparse-input} and~\ref{thm:gap-circuit-input} that shows hardness of $\alpha$-approximate $\PolySparse_R$ for some small $\alpha$. Then we shall amplify this hardness for any $\beta >1$.
\begin{theorem}\label{thm:small-gap}
  Let $R = \F_p, 
  \mathbb{R}, \mathbb{Q}$ or $\mathbb{Z}_q$. For all $\varepsilon, \delta >0$ as given in the last column of  \cref{table:max3lin-hardness}, there exists an $\alpha = \alpha(\varepsilon, \delta, R)$ such that it is $\NP$-hard to $\alpha$-approximate $\PolySparse_R$, in either the sparse, dense or arithmetic circuit representation.
\end{theorem}
\begin{proof}
  
  We first note that as the input is a degree $2$ polynomial, all three representations are polynomially equivalent.
  
  Suppose for a regime of values of $\varepsilon', \delta' >0$ we are guaranteed the following. Given an instance of $\maxthreelin_R$, it is $\NP$-hard to distinguish the following cases -- if there is a assignment that satisfies at least $(1-\varepsilon')$ fraction of linear equations or for all assignments at most $\delta'$ fraction of linear equations are satisfied. Putting this together with \cref{lem:gap-reduction}, we get that it is $\NP$-hard to distinguish if there is a vector $\veca$ such that the polynomial $Q_S(Y+\veca)$ has at most $t = 3m+\varepsilon' m+1 $ monomials or if for all $\veca$, the polynomial $Q_S(Y+\veca)$ has at least $\alpha t = (4-\delta')m$ non-constant monomials. Thus, we get that it is $\NP$-hard to $\alpha$-approximate $\PolySparse_R$ where $\alpha = \alpha(\varepsilon', \delta')$ is obtained as follows.
  \begin{align*}
    \alpha = \frac{4-\delta'}{3+\varepsilon' +\frac{1}{m}} = \frac{4}{3} - \frac{4\varepsilon' +3\delta'+o(1)}{9+3\varepsilon'+o(1)}\,.
  \end{align*}
  Each row of \cref{table:max3lin-hardness} gives us a guarantee of the form that we assumed at the beginning of this proof. Thus by iterating through the rows of \cref{table:max3lin-hardness}, we get our parameter $\alpha = \alpha(\varepsilon, \delta, R)$ for various settings of $R$.  This completes the proof.
\end{proof}

Let $d$ be a parameter that we shall soon fix. Let $Y^{(1)}, \ldots, Y^{(d)}$ be $d$ many disjoint copies of the variable set $Y = \{y_1, \ldots, y_w\}$. Let $Y_d = \sqcup_{k=1}^d Y^{(i)}$. Let the polynomial $F_{n,d}(Y_d)$ be defined as follows.
\begin{align}\label{eqn:amplification-lin}
    F_{n,d}(Y_d) = \prod_{k=1}^d Q_S(Y^{(k)}).
\end{align}
Observe that the sparsity of $F_{n,d}(Y_d)$ is given by the product of sparsities of $d$ many instances of $Q_S(Y)$. Further, if the polynomial $Q_S(Y)$ is computed by a circuit of size $s$ then the polynomial $F_{n,d}(Y_d)$ has a circuit of size at most $sd+1$.

\begin{lemma} \label{lem:gap-amplification-large}
  Let $S$ be a system of linear equations, and $F_{n,d}(Y_d)$ be the polynomial as defined above. 
  \begin{enumerate}
  \item All vectors $\veca\in R^n$ satisfy at most $\delta$ fraction of equations in $S$ if and only if  all vectors $\vecb_d\in R^{wd}$ are such that $F_{n,d}(Y_d+\vecb_d)$ has at least $((4-\delta)m)^d$ non-constant monomials.\label{item:4-d}
  \item There exists a vector $\veca\in R^n$ such that it satisfies at least $1-\varepsilon$ fraction of equations in $S$ if and only if there exists a vector $\vecb_d\in R^{wd}$ such that $F_{n,d}(Y_d+\vecb_d)$ has at most $((3+\varepsilon)m+1)^d$ monomials.\label{item:3+e}
  \end{enumerate}
\end{lemma}

\begin{proof}
From the product structure of $F_{n,d}(Y_d)$, we get that for  all vectors $\vecb_d\in R^{wd}$, the polynomial $F_{n,d}(Y_d+\vecb_d)$ has at least $((4-\delta)m)^d$ non-constant monomials if and only if for all vectors $\veca\in R^n$, the polynomial $Q_S(Y+\veca)$ has at least $(4-\delta)m$ non-constant monomials. For the sake of contradiction, let us suppose that there is a vector $\vecb'_d\in R^{wd}$ such that $F_{n,d}(Y_d+\vecb'_d)$ has at most $((4-\delta)m)^d-1$ monomials. Because of the product structure of $F_{n,d}(Y_d)$, it must the case that there is a copy of $Q_S(Y)$, say $Q_S(Y^{(i)})$ such that $Q_S(Y^{(i)}+\vecb'_d|_{Y^{(i)}})$ has at most $(4-\delta)m-1$ non-constant monomials which contradicts our assumption. The other direction also follows trivially from the product structure. From \cref{lem:gap-reduction}, we get that $\veca'$ satisfies at most $\delta$ fraction of equations in $S$ if and only if $Q_S(Y+\veca)$ has at least $(4-\delta)m$ non-constant monomials. By putting both of these together, we get \cref{item:4-d}.

By invoking \cref{lem:gap-reduction} again, we get that  there is a vector $\veca\in R^n$ such that it satisfies at least $1-\varepsilon$ fraction of equations in $S$ if and only if there is a vector $\vecb$ such that $Q_S(Y+\vecb)$ has at most $(3+\varepsilon)m+1$  monomials. By taking $\vecb_d$ to be the concatenation of $\vecb$, $d$ many times, we get that $F_{n,d}(Y_d+\vecb_d)$ has at most $((3+\varepsilon)m+1)^d$ monomials. On the other hand, if there exists a vector $\vecb_d\in R^{wd}$ such that $F_{n,d}(Y_d+\vecb_d)$ has at most $((3+\varepsilon)m+1)^d$ monomials then because of the product structure of $F_{n,d}(Y_d)$, $Q_S(Y+\veca)$ has at most $(3+\varepsilon)m+1$  monomials where $\veca = \vecb_d|_{Y^{(1)}}$. This completes the proof of \cref{item:3+e}.
\end{proof}

\begin{proof}[Proof of Theorems~\ref{thm:gap-sparse-input} and~\ref{thm:gap-circuit-input}]
Given any $\beta>1$ and $\alpha$ as given by \cref{thm:small-gap}, let $d = \log_{\alpha}{\beta}$. Thus, $\beta = \alpha^d$. Let $F_{n,d}(Y_d)$ be the polynomial as defined in \cref{eqn:amplification-lin}. From \cref{lem:gap-amplification-large}, we get that $\alpha$-gap-$\PolySparse_R$ gap reduces to $\alpha^d$-gap-$\PolySparse_R$.
  
  To prove Theorem~\ref{thm:gap-sparse-input} we note that if $Q_S(Y)$ is provided in sparse representation (recall that sparsity of $Q_S(Y)$ , denoted by $t$, is at most $4m+1$) and if $\alpha^d$-gap-$\PolySparse_R$ problem for $F_{n,d}(Y_d)$ can be solved efficiently in time $N^{O(1)}$ (where $N=t^d$ is the sparsity of the polynomial $F_{n,d}(Y_d)$) then $\alpha$-gap-$\PolySparse_R$ $Q_S(Y)$ can be solved in time $t^{O(d)}$. Thus, as long as $d=O(1)$ the gap reduction runs in polynomial time.
  
  Similarly, to prove  Theorem~\ref{thm:gap-circuit-input} we note that if $Q_S(Y)$ is provided as a circuit of size $s$ and if $\alpha^d$-gap-$\PolySparse_R$ problem for $F_{n,d}(Y_d)$ can be solved efficiently in time $N^{O(1)}$ (where $N=sd+1$ is the input size of the instance provided as a circuit) then $\alpha$-gap-$\PolySparse_R$ problem $Q_S(Y)$ can be solved in time $\inparen{sd}^{O(1)}$. As long as $d$ is at most a polynomial in $s$, the gap reduction runs in polynomial time.
\end{proof}

\bibliographystyle{alphaurl}
\bibliography{ref}

\end{document}